\documentclass[9pt,twocolumn,twoside]{osajnl}
\usepackage{xcolor}
\usepackage{amsmath}
\usepackage{amsmath}
    \usepackage{amssymb}
    \usepackage{amsthm}
    \usepackage{amsfonts}
    \usepackage{braket}

\journal{ol} 

\setboolean{shortarticle}{true}

\title{Quantum memories and the double-slit experiment: implications for astronomical interferometry}

\author[1,2,3]{Joss Bland-Hawthorn}
\author[4]{Matthew J. Sellars}
\author[1,5,6]{John G. Bartholomew}

\affil[1]{School of Physics, University of Sydney, NSW 2006, Australia}
\affil[2]{Sydney Astrophotonic Instrumentation Labs (SAIL), School of Physics, University of Sydney, NSW 2006, Australia}
\affil[3]{Sydney Institute for Astronomy (SIfA), School of Physics, University of Sydney, NSW 2006, Australia}
\affil[4]{Centre for Quantum Computation and Communication Technology, Research School of Physics and Engineering, The Australian National University, Canberra 0200, Australia}
\affil[5]{Centre for Engineered Quantum Systems, School of Physics, The University of Sydney, Sydney, NSW 2006, Australia}
\affil[6]{The University of Sydney Nano Institute, The University of Sydney, NSW 2006, Australia
}

\affil[*]{Corresponding author: jbh@physics.usyd.edu.au}

\begin{abstract}
Thomas Young's slit experiment lies at the heart of classical interference and quantum mechanics.
Over the last fifty years, it has been shown that particles (e.g. photons, electrons, large molecules), even {\it individual} particles, generate an interference pattern at a distant screen after passage through a double slit, thereby demonstrating wave-particle duality. 
We revisit this famous experiment by replacing both slits with single-mode fibre inputs to two independent quantum memories that are capable of storing the incident electromagnetic field's amplitude and phase as a function of time. At a later time, the action is reversed: the quantum memories are read out in synchrony and the single-mode fibre outputs are allowed to interact consistent with the original observation. In contrast to any classical memory device, the write and read processes of a quantum memory are non-destructive and hence, preserve the photonic quantum states.
In principle, with sufficiently long storage times and sufficiently high photonic storage capacity, quantum memories operating at widely separated telescopes can be brought together to achieve optical interferometry over arbitrarily long baselines.
\end{abstract}

\setboolean{displaycopyright}{true}

\begin{document}

\maketitle

\section{Introduction}

\subsection{Very long baseline interferometry}

In April 2019, a worldwide telescope network called the Event Horizon Telescope (EHT) presented their discovery of the shadow of a supermassive black hole \cite{EHT2019}. 
Eight radio telescopes at six different geographic locations around the Earth were used to observe the centre of the giant elliptical galaxy M87 in the Virgo cluster. The observations were carried out at $\lambda =$~1.3~mm (230 GHz) where the data stream was time-stamped using a local atomic clock at each node. The data were stored on conventional hard drives and then flown to a common location where the separate data streams were synchronized to within 10$^{-12}$ s before being correlated. The final image, with its angular resolution of 20 $\mu$as (0.1 nano-radians), was a remarkable demonstration of the power of radio-frequency ``very long baseline interferometry'' (VLBI).

Optical interferometry preceded radio interferometry by more than a century. After Young's slit experiment in 1803, early attempts to demonstrate optical interference over metre-scale baselines culminated in Michelson and Morley's famous 1887 demonstration of the finite speed of light for all reference frames. Today, optical interferometers preserve the light's phase information over baselines as long as 300 m \cite{Monnier2003}, a distance that is orders of magnitude smaller than what is possible at radio wavelengths. The primary challenge is the limited surface brightness of optical sources compared to their radio counterparts. The elevated brightness temperatures of compact sources, e.g. M87's core regions ($\sim 10^{9-10}$ K), contributes to the success of radio VLBI.

\begin{figure}[htbp]
\centering
\fbox{\includegraphics[width=\linewidth]{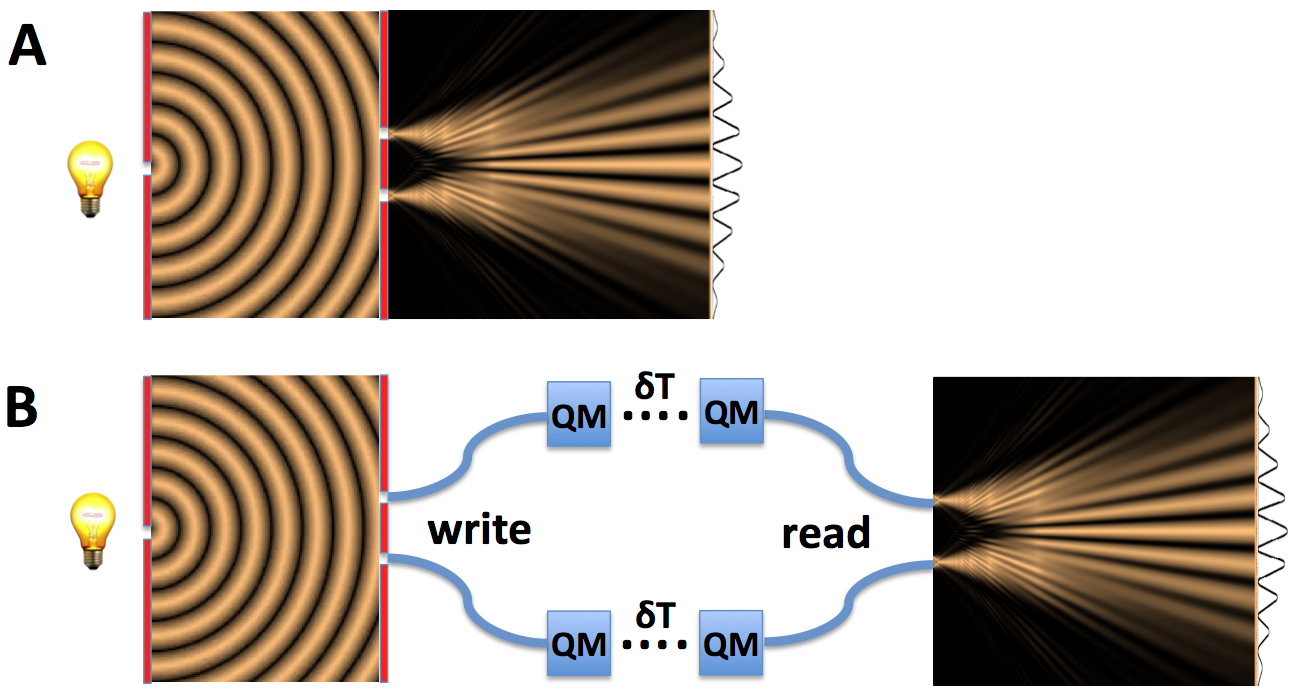}}
\caption{(A) The conventional Young's double slit experiment illustrating how a coherent source generates an interferogram at a distant screen, assuming the slits are diffraction limited. The line trace is the amplitude of the electric field. (B) To illustrate our new concept, the same experiment is now split into two parts. Two single-mode fibres act as the double slit and direct photonic states to be written into independent quantum memories (QMs). The ellipses represent time passing ($\delta$T), after which the photonic data stored in the QMs are read out into single-mode fibres and their well-aligned outputs are allowed to interfere. The interferogram is recovered for which every detected photon arises from a superposition of information stored in {\it both} QMs.
}
\label{f:QMslits}
\end{figure}

There is a scientific case for achieving longer baselines at optical wavelengths. Sub-microarcsecond resolution (below the current optical limit) would allow some stellar types to be imaged, along with exotic sources closer to home (e.g. white dwarfs). In principle, it would allow the distances to millions of galaxies to be determined from their astrometric motions or parallax motions as the Earth orbits the Sun. 
These futuristic statements are not defended here, but we identify prohibitive photon loss over long baselines as one of the key challenges preventing optical VLBI today, and propose a solution. 

The prospect of an optical VLBI has long been discussed but considered beyond reach at the present time \cite{Dravins2005}. But now, we see a way forward. The international race to build a quantum computer (QC) encompasses significant efforts to create quantum networks to faithfully distribute quantum states (amplitude and phase information), potentially over long distances. Such networks will be essential if the different QC nodes are to communicate effectively (e.g. for distributed quantum calculations). To underscore the difficulty, consider this statement from quantum information theory. The {\it no-teleportation theorem} states that an arbitrary quantum state cannot be converted into a (finite or infinite) sequence of classical bits, nor can such bits be used to reconstruct the original state. Thus, teleporting a state by merely moving classical bits around does not work because the unit of quantum information, the {\tt qubit}, cannot be precisely converted into classical information bits because of the Heisenberg uncertainty principle. Equivalently, the amplitude and phase of a photon cannot be measured, recorded and then recreated, nor can a photon simply be amplified to overcome the loss in a distribution channel. An entirely new approach is needed.

\subsection{A new experiment}

Consider the two experiments illustrated in Fig.~\ref{f:QMslits}. The upper apparatus in (A) is the conventional Young's double-slit experiment where a diffraction limited source (e.g. laser) or any source illuminating a diffraction-limited slit illuminates two slits at the next screen. 
In the limit of low illumination (the single particle regime), if an attempt is made to measure which slit the particle passed through, the interference pattern is lost, in agreement with Heisenberg's uncertainty principle. 

We now propose an equivalent experiment in (B) that has profound implications for astronomical interferometry and related fields in physics. The two slits are replaced with independent storage systems (QMs) that are able to preserve input photonic quantum states (the phase and amplitude of light as a function of time). At a later time, if the QMs can be read out in synchrony to faithfully recall the stored photons the original interference pattern is recovered.
The write and read operations in Fig.~\ref{f:QMslits} can be executed {\it without measurement} allowing the photonic quantum state to be preserved. 

\begin{figure}[htbp]
\centering
\fbox{\includegraphics[width=\linewidth]{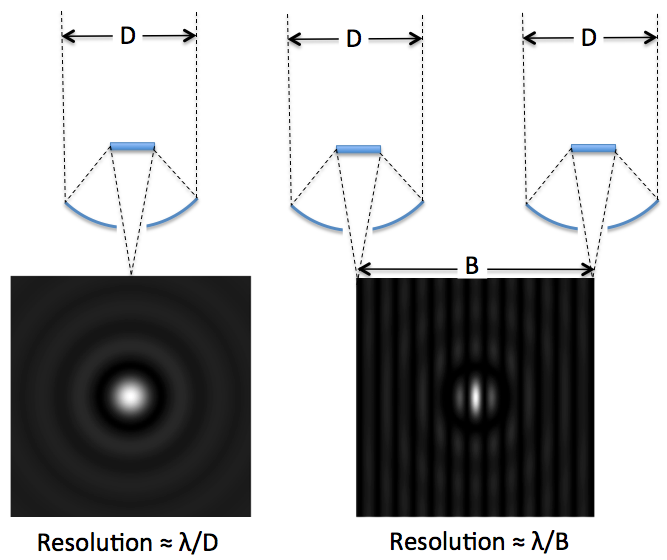}}
\caption{When QMs are employed at three or more distributed identical telescopes, it is possible to reconstruct (Left) the diffraction-limited response of a single telescope with diameter D; (Right) the diffraction-limited response of each baseline, B, where B$\gg$D.
}
\label{f:Baseline}
\end{figure}

In~Fig.~\ref{f:Baseline}, we illustrate how two identical, diffraction-limited telescopes (with diameter $D$) in a multi-node interferometer (e.g. EHT experiment) achieve higher spatial resolution. When the data are collected and assembled at one location for later processing, the original baseline separation $B$ can be replaced with an equivalent time delay to preserve phase information when the stored data are clocked out and allowed to interfere. The importance of the set-up in (B) is that the storage devices can be placed far apart and then brought together if the storage time is sufficiently long. 

While there are ways to trap mass-carrying particles indefinitely (e.g. laser cooling), stopping a photon in its tracks presents a more difficult problem. Indeed, in astronomy, all successful methods to date are based on trapping a photon (strictly, an energy packet) in an optical fibre coil or a high-$Q$ optical cavity. The best telecom fibres have 0.142 dB/km loss such that $>$60\% of photons are lost after propagating 30 km (storage time $\sim$ 0.1 ms). The best optical cavities allow up to a million or more internal reflections before photons are absorbed or scattered. In both instances, the trapping time is short ($\ll 1$ sec) before the photon is eventually lost to the system. Given the importance of quantum memories to the idea of long distance quantum networks, many researchers have begun to consider new ways to solve the photon trapping problem. 

It is important to recognize that the information carried by the photonic quantum state encoded by the energy packet, is what matters. We define a quantum memory for light as a static device where the fragile quantum state of a photon can be faithfully and efficiently stored and recalled at a later time. Thus, given that particles with mass can be trapped indefinitely, a remarkable prospect is to map the photonic state into the quantum state of an atomic system.

So what kind of light-matter interaction allows for a photon's quantum state to be transferred to an atomic state? Light-matter interactions are typically very weak: to achieve a high probability of the photonic information being written onto a single atom, we first consider an energy packet strongly interacting with an atom embedded within a high-$Q$ optical cavity\footnote{This field, collectively known as cavity quantum electrodynamics (CQED), was recognized by the 2012 Nobel Prize in Physics awarded to Serge Haroche and David Wineland. For a review, we refer the reader to \cite{Walther2006}.}. The atom must then be safeguarded against its noisy environment during storage because the fragile quantum state can be easily disrupted. 
The work in Ref.~\cite{Specht2011} realized a read-write QM using a single Rb atom trapped within an optical cavity. But the first demonstration of a QM for light with an efficiency greater than 50\% was enabled by mapping photons onto the collective state of a large ensemble of atoms `trapped' in a solid-state crystal~\cite{Hedges2010}. Here the strong light-matter interaction was achieved by using $10^{13}$ atoms, overcoming the very weak interaction at a single atom level. We discuss a proposal that combines elements of these two approaches - coupling ensembles of atoms in the solid-state to high-$Q$ photonic cavities - that allows for the storage of large numbers of photons and for the memory to be transported, while demonstrating the potential for storage times that are interesting for practical applications.


\section{Distributing quantum states}

For a quantum network, as for an optical VLBI, the goal is to bring the observed, delicate quantum states (probability amplitudes) from widely separated locations together in order to generate an interference pattern. We briefly consider some of the ideas under discussion, some old, some new.

\subsection{Optical fibres vs. free space}

As noted by \cite{Loffler2011}, ``conventional multimode fibres suffer from strong intermodal coupling that tends to destroy the fragile quantum correlations carried by the spatially entangled state.'' There are optical fibre designs that preserve multiple transverse modes of the 
propagating field. For example, Ref.~\cite{Loffler2011} investigated the properties of a hollow-core
photonic crystal fibre (PCF) with kagome-style cladding, core diameter 25~$\mu$m and length 30~cm.
Recent developments suggest a simpler structure
can be made with the same propagational properties but both fibres are lossy and prohibitively expensive over long distances. Significantly lower loss is possible in SMF-28 or few-mode fibre, allowing demonstrations of photonic entanglement distribution over 200 km distances~\cite{Dynes2009,Cui2017}. But even then, the best telecom products have 0.14 dB/km absorption loss such that approximately 30\% of light is lost after 10 km, and $>$95\% is lost after a distance of 100 km, at best. 

In free-space propagation, if we ignore the impact of a turbulent Earth's atmosphere, the photonic loss due to absorption is many orders of magnitude less and quantum entanglement distribution links have been established over great distances ($\sim$1200~km) between an Earth station and a space satellite~\cite{Chen2021}. But free-space propagation is currently limited by inverse-square dilution, for example 22~dB for a 1200~km link~\cite{Liao2017}.

\subsection{Quantum repeaters}

Establishing a long-distance multi-node quantum network is an active area of research both experimentally and in quantum
information theory. As discussed in the previous section, direct quantum links based on distributing photons through optical fibres suffer losses that scale exponentially with distance due to scatter, diffraction and absorption during transmission. This loss needs to be overcome for fibre channels to be scalable: that is, not requiring exponentially increasing resources to add distance or nodes to the network. 

In classical telecom networks, optical repeaters are used to boost the signal with the aid of optical amplifiers. In contrast, there is no way to amplify the signal of a quantum state because it is impossible to make multiple identical copies of an unknown quantum state without altering the original state $-$ i.e. there is no quantum cloning\footnote{Interestingly, imperfect copies can be made, a concept known as quantum copying \cite{Buzek1996}; this concept is potentially useful in quantum interferometry with regard to maintaining the metrology and stability of the system.} $-$
a new approach is called for. 

Quantum repeaters are one protocol that combat exponential fibre losses, and are
designed to allow arbitrarily long-distance entanglement distribution with losses that scale polynomially with distance. 
The original quantum repeater scheme was proposed by \cite{Briegel1998}. The concept is to divide the total transmission length of the network into $2n$ shorter lengths over which fibre transmission loss is negligible. The segmented transmission channel is punctuated by $2n+1$ quantum nodes that will enable the operations on which the protocol relies. An entangled photon pair source is stationed at every $2i$th node. From each source, one photon is sent to node $2i - 1$ and the other sent to $2i + 1$. At $2i + 1$th nodes, quantum memories (\S 3) are placed to receive each transmitted photon. If the photon arrives, the memory stores that mode. If the photon is lost, the source is triggered again until both photons reach their destination and are stored. An entanglement swapping procedure (Bell state measurement) is then performed on the pairs of quantum memories at each of the relevant intermediate nodes, which entangles the states of the quantum memories at the start and end point of the transmission length.

Quantum repeaters are still in their infancy and require breakthroughs across a wide array of technological frontiers, including single photon sources, entangled photon pair sources, quantum memories, Bell state measurement circuits, and so forth. Our new approach focuses on just one of these technologies: the quantum memory. However, in contrast with the repeater protocol, rather than fixing the quantum memory in space, we propose to transport the memory during the photon storage, which is often referred to as a quantum hard drive\footnote{Entanglement distribution by distributing quantum hard drives has previously been referred to as a quantum sneakernet~\cite{Devitt2016}.}.

\subsection{Quantum hard drives}

Conventional computers distinguish between random access memory (temporary storage) and the hard drive (long-term storage). Quantum random access memory (QRAM) is a key aspect of quantum computer architecture and has very different constraints compared to those required by an interferometer.
The more interesting concept for our application
is the quantum hard drive (QHD) that is able to 
preserve quantum information with high efficacy for long timescales compared to the time taken to physically transport them - ideally days or even years. While we are unaware of a physical system that achieves the latter, the former is feasible given recent developments. If we were able to store quantum information at nodes A and B on QHDs, these devices could
then be transported to intermediate node C (see Fig.~\ref{f:jukebox}.a). The stored signals on a local QHD would then be
read out and combined with the other QHD, or QHDs if this is a multi-node network.

The QHD is an alternative protocol to the quantum repeater that could in principle achieve low-loss optical mode transport over long distances. The first prospect for a practical implementation of such a device was in Ref.~\cite{Zhong2015} who demonstrated coherence lifetimes of optically addressable nuclear spin states in Eu$^{3+}$-doped crystals to be in excess of 6 hours. At this low decoherence rate, transporting a QHD at speeds $>$10 km/hour would achieve lower losses during transport than light propagating in the lowest loss optical fibre (the threshold speed reduces to less than 0.5 km/hour at the relevant wavelength for Eu$^{3+}$: 580~nm). With numerous optical QM demonstrations in the same material~\cite{Jobez2014, Jobez2015, Jobez2016, Laplane2016, Laplane2017, Ma2021} and excellent prospects for extending the spin-state coherence lifetime beyond 1~day~\cite{Zhong2015}, it is a promising technology. If high efficiency, high fidelity and high data storage capacity can be achieved, photonic quantum states gathered at different nodes can be brought together from $\sim 100$~km distances. The long-term goal is to achieve lengthier storage times and reliable transportability over continental distances.

How well a QHD can perform is still an open questions (see \S~\ref{s:challenges}\ref{subs:transport}), for example, the transportability of QMs has yet to be demonstrated over any significant distance.  
We now take a closer look at QM technology that lies at the heart of our new scheme and how it is to be harnessed in an interferometer.

\section{Quantum memory operations}

\subsection{Storing photonic states in an atomic memory: concepts}

A photon encodes four parameters: these are orbital angular momentum and 3 components of the linear momentum vector. Equivalently, when the wave-packet is confined to a beam, it is usefully described in terms of polarisation, the spatial mode profile (two degrees of freedom), and energy. An ideal QHD to enable optical VLBI would faithfully store the complete quantum state of the incident photon and that storage process would not add any noise upon reading the stored signal. This is indeed possible with the QHD we consider in this paper because the operation is analogous to a four dimensional holographic medium that stores the amplitude and phase of input light as a function of time. As discussed in \S~3\ref{s:density}, long-term storage of phase information currently sets stringent bounds in terms of memory bandwidth. In contrast, storage of photonic intensity as a function of time in a similar device relaxes those bounds significantly. This suggests that Hanbury-Brown-Twiss measurements may prove to be a beneficial stepping stone toward the the more difficult challenge of phase interferometry that we consider in this work.

Consider an atomic ensemble memory where all the atoms are initialised into a specific spin-energy level in their optical ground state  $\vert g \rangle$ (see Fig.~\ref{f:multiplexing}). A {\it single} photon input\footnote{Throughout this paper, the use of ``photon'' is shorthand for an energetic wave-packet described by a photonic quantum state.} into the ensemble that is resonant with an optical transition $|g\rangle~\Longleftrightarrow~|e\rangle$ to an optical excited-state energy level $|e\rangle$ will be absorbed. Even though each atom only interacts weakly with the photon (Chanali\`ere et al 2005~\cite{Chaneliere2005}; Eisaman et al 2005~\cite{Eisaman2005}), the collective optical depth of the ensemble or an atomic-cavity system can approach perfect absorption. The absorption event creates a single atomic excitation that is spread out across the entire spatial and frequency extent of the ensemble $-$ a so-called Dicke state $-$ that has a time evolution

\begin{equation}
    |\psi(t)\rangle = \sum_{j = 1}^N \alpha(\boldsymbol{r}_j, \delta_j) e^{-i(\boldsymbol{k\cdot}\boldsymbol{r}_j - \delta_j t) -\beta t} |g_1 \cdot\cdot\cdot e_j \cdot\cdot\cdot g_N\rangle~,  
\end{equation}
where $\boldsymbol{r}_j$ is the position of the atom, $\boldsymbol{k}$ is the photonic field wavevector, $\delta_j$ is the detuning of the atom with respect to the photon frequency, $\beta$ is the atomic decoherence rate, and the amplitude $\alpha$ is a function of the atom's position and frequency.

Another way of considering the Dicke state is that the lack of knowledge as to which specific atom absorbed the photon results in a large entangled state, which is a sum of all the possible combinations of only one out of the billions of atoms being excited.  

At $t = 0$, all the atomic dipoles are initially phase-aligned with $\boldsymbol{k}$ but the ensemble rapidly dephases because of inhomogeneity: each atom accumulates phase at a different rate $\delta_j$. Remarkably, this is a reversible process. In this work we consider a photon-echo quantum memory protocol (see \S~3\ref{s:density}), which is an optical analogy of spin echo techniques in nuclear or electron magnetic resonance. The memory protocol acts to rephase the atomic dipoles at a later time $t = t^\prime$ to create the ensemble coherent state

\begin{equation}
    |\psi(t^\prime)\rangle = \sum_{j = 1}^N \alpha(\boldsymbol{r}_j, \delta_j) e^{-i(\boldsymbol{k\cdot}\boldsymbol{r}_j) - \beta t^\prime} |g_1 \cdot\cdot\cdot e_j \cdot\cdot\cdot g_N\rangle~.  
\end{equation}

This recreates the collective coherence of the absorption event, that is $|\psi(t^\prime)\rangle = e^{-\beta t^\prime}|\psi(0)\rangle$, and the in-phase dipole array re-emits the absorbed photon, preserving the phase and spatial information of the input state. In the limit $t \ll \beta^{-1}$, the loss due to decoherence can be negligible and stored quantum states can be recalled with an efficiency approaching unity, although practical challenges have limited current implementations to efficiencies less than 90\% \cite{Hedges2010, Cho2016, Wang2019}. The description above is restricted to a single photon stored by a photon-echo quantum memory, but the protocol can faithfully store \emph{any} weak input field that is within the memory bandwidth. For example, the memory operation can also be described using a Maxwell-Bloch formalism for weak time-varying input fields~\cite{Longdell2008}. Thus, upon readout the full time-sequence of input photonic states is reconstructed. There are also numerous protocols that realise a quantum memory beyond what we describe in this work. In the interest of brevity, we refer the reader to several introductions of how the ensemble read-write (capture and release) QM system is achieved in practice~\cite{Lvovsky2009,Afzelius2015,Heshami2016,Campbell2016,Chaneliere2018}.

\subsection{Long-term storage and transport}
The storage time $t$ that allows efficient recall of information is capped by the decoherence loss term $e^{-\beta t}$. For optical transitions, the decoherence rate $\beta$ often approaches the lower bound of
\begin{equation}
\beta = \dfrac{1}{T_{1, \rm o}}~,
\end{equation}
where $T_{1, \rm o}$ is the optical lifetime of state $|e\rangle$. This sets the efficient storage time limit $t \ll T_{1, \rm o}$. 

To extend beyond the regime of optical lifetimes the photonic state encoded on the optical transition $|g\rangle~\Longleftrightarrow~|e\rangle$ can be transferred to a longer lived spin transition $|g\rangle~\Longleftrightarrow~|s\rangle$ (see Fig.~\ref{f:multiplexing}). The decoherence rate $\beta$ for spin transitions is typically significantly less than the lifetime limit $1/T_{1, \rm s}$ due to perturbations caused by magnetic or electric field fluctuations. It is the \emph{coherence lifetime} of the spin transition - $T_{2, \rm s}$ - that set the upper bound on the quantum storage time ($t \ll T_{2, \rm s}/2$).
For the majority of atomic systems under investigation for photonic QMs the storage time has been limited to much less than 1 second~\cite{Cho2016}. While there have been examples of light storage for around 1~minute~\cite{Heinze2013, Dudin2013}, this is still insufficient for optical VLBI. 

Here we highlight one particular system that possesses the key requirements for a quantum hard drive for optical VLBI: Eu$^{3+}$-doped Y$_2$SiO$_5$. The frozen ensemble of europium atoms couple weakly to the lattice at liquid helium operating temperatures ($<$4~K) but can have high spectral density allowing photons to be absorbed effectively. The Eu$^{3+}$ $^7$F$_0 \leftrightarrow ^5$D$_0$ optical transition also possesses a homogeneous linewidth of $<200$~Hz~\cite{Equall1994}, and the nuclear spin $I = 5/2$ exhibits clock transitions that have first order insensitivity to magnetic field perturbations.

In 2015, Zhong et al. demonstrated a nuclear spin coherence lifetime of $T_{2, \rm s} > 6$ hours in a specifically oriented applied magnetic field, {\it thereby demonstrating the potential of long-lived photonic states}~\cite{Zhong2015}. The use of dynamical decoupling, pulses applied to the nuclear spin every 100~ms analogously to techniques long employed in nuclear magnetic resonance, cancels out residual magnetic noise and ensures robustness and long-term storage. Excitingly, the successful mapping and recall of photonic information stored on the optical transition to these long-lived nuclear spin states has recently been demonstrated~\cite{Ma2021}.
In principle, the quantum storage time is bounded by the lifetime of the nuclear spin to 46 days limited by spin-lattice relaxation~\cite{Konz2003}, but magnetic perturbations from the  yttrium nuclear spins in the host set the current bound of hours. Furthermore, a quantum memory that can travel (QHD) has also yet to be realized. Both long storage times and the ability to preserve the quantum information during transport are central to our proposed development.

\begin{figure*}[htbp]
\centering
\fbox{\includegraphics[trim=0 70 0 0, clip,width=0.9\textwidth]{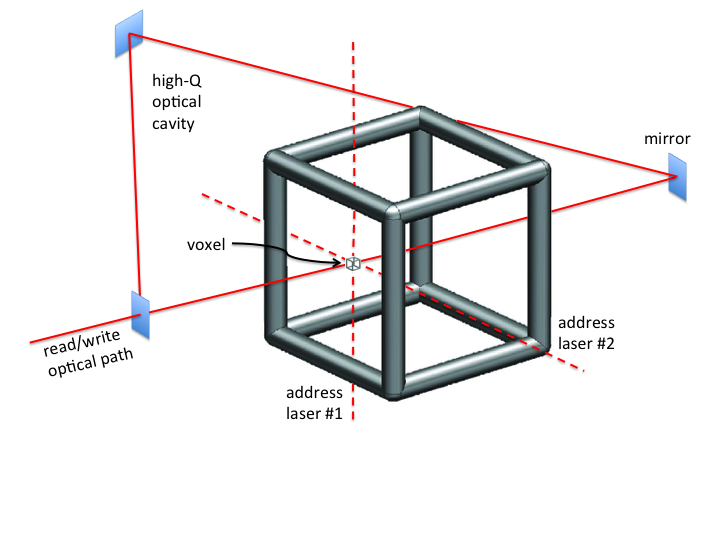}}
\caption{Schematic diagram of a Eu$^{3+}$:YSO crystal coupled to a high-Q optical cavity. A specific voxel can be addressed for the read/write operation with the aid of one or more address lasers (dashed lines). Direct write of a photon's quantum state into a voxel with a single laser path is possible (solid line), but the operation is far more efficient if the voxel resides within an optical cavity (shown by the triangle) at the time of the read/write operation. Two mirrors are highly reflective but the third mirror on the read/write path is partially reflective; this mirror admits the photon if it couples to the cavity mode.}
\label{f:cavity}
\end{figure*}

\begin{figure*}[htbp]
\centering
\fbox{\includegraphics[width = 0.9\textwidth]{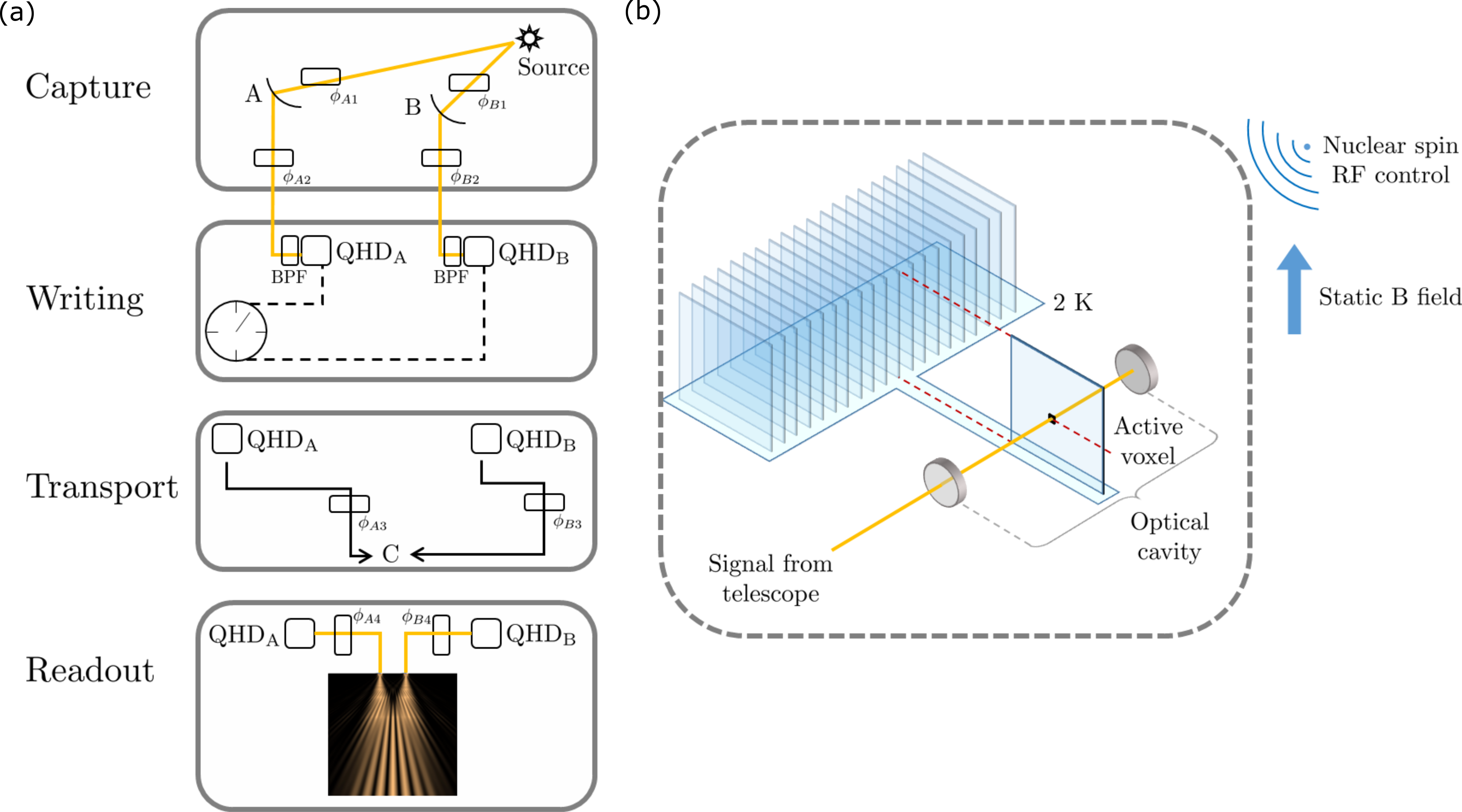}}
\caption{Optical VLBI enabled by QHDs. (a) Two terrestrial telescopes separated by some
baseline collect light from a source (Fig.~\ref{f:Baseline}), with the $\phi_{Ai}$ and $\phi_{Bi}$ representing path dependent phase shifts during the different processes. Following the signal capture and routing into single mode fibres at each telescope, the signal is prefiltered with a series of narrow spectral bandpass filters (BPFs) and written onto a QHD (see text and Fig.~\ref{f:multiplexing} for further details). Accurate clocking is needed to synchronise the QHD exposure periods and to track their relative phase evolution. Once the writing process is completed, the QHDs are transported (feasibly of the order of a few hundred kilometres) to a third site for data processing. The preserved photonic signals are readout and combined, in this case to produce a spatial interferogram. (b) The conceptual framework for the QHD based on long-term storage of light on the internal energy levels of atoms embedded in crystals. A jukebox mechanism sequentially loads disks into an optical cavity. The signal from the telescope is coupled to the cavity allowing the photons to be absorbed by the atomic ensemble in the active voxel through a collective coherent interaction. The active voxel is controlled by rastering the cavity waist relative to the disk, allowing a large number of photons to be stored across the spatial degree of freedom. The 20 disk jukebox as shown contains $2\times10^7$~voxels and could store up to $2\times10^{10}$ optical modes under our current assumptions. To enable the data to be written, stored, transported, and read out, the QHD must be maintained at a stable temperature of $\sim2$~K, with a precisely aligned orientation relative to a static external magnetic field, and be controlled using pulsed radio frequency signals. The pictured configuration would have a polarisation-dependent storage efficiency because absorption is anisotropic in Eu$^{3+}$:YSO. A simple extension that makes use of two perpendicular crystal orientations in a single disk~\cite{Clausen2012} would allow storage of arbitary polarisation states.}
\label{f:jukebox}
\end{figure*}

\subsection{Increasing the photonic storage density: concepts}
\label{s:density}

Our goal here is to provide an approach to writing vast numbers of photons into a quantum memory with a view to storing an entire astronomical observation. Several groups around the world have managed to store multiple photonic states at a quantum level using a combination of temporal, frequency, and spatial degrees of freedom~\cite{Yang2018} but we present a method for highly multiplexed long-term storage making use of all three properties.

To make the QHD practical, the photonic storage density needs to be high enough to overcome the latency of the distribution scheme while ensuring the QHD does not become too large. We first consider the spatial degree of freedom and ask the question, how small could we make a single storage voxel (see Fig.~\ref{f:cavity})? The voxel size is limited by a combination of the absorption of the atoms within the voxel and the spatial mode of the incoming photons. The Eu$^{3+}$ concentrations used to demonstrate long-term storage in Y$_2$SiO$_5$ crystals~\cite{Zhong2015,Ma2021} are typically at impurity levels (${\cal C} \approx 10^{-4}-10^{-3}$ relative to Y). The optical transition of Eu$^{3+}$:YSO is very weak such that at these concentrations the maximum absorption is of the order of 4~cm$^{-1}$~\cite{Konz2003}. In principle, the concentration can be much higher, for example stoichiometric crystals with ${\cal C} \approx 1$ that yield peak absorption coefficients as high as 1000~cm$^{-1}$~\cite{Ahlefeldt2013a, Ahlefeldt2016}, but their use in long storage time QMs has yet to be demonstrated.

Irrespective of the underlying optical depth of the atomic medium, to minimise the size of each spatial voxel, storage protocols heavily benefit from the coupling of the atoms in the voxel to the mode of a high-Q resonator~\cite{Afzelius2010}. By matching the optical absorption per cavity round trip to the transmission of the coupling mirror - the ``impedance matching'' condition - photons coupled to the cavity can in principle be perfectly absorbed by the ensemble. This configuration is illustrated in Fig.~\ref{f:cavity} where each voxel resides within a 3-mirror cavity. The ``read/write optical path'' is where the incoming/outgoing photon meets a partially reflecting mirror; the other mirrors have very high reflectivity. An incoming photon, for example, is accepted by the cavity if it couples to the cavity mode. This is a standard problem in CQED that is closely related to many mode-coupling problems in astrophotonics.

For the present work, we set our memory voxel to be $60\times 60\times 100$ $\mu\rm m^3$. Although the single-pass optical depth is low, approximately 0.04, high efficiency storage is feasible so long as round trip losses in the cavity are much less~\cite{Jobez2014}. For example, the impedance matching condition could be achieved in a $\sim$20~cm long Fabry-Perot cavity with a finesse of 80. The typical maximum commercially available cross-sectional diameter of a YSO crystal is ${\cal L}_{xy} \approx 60-80$~mm where its length can be up to ${\cal L}_z \approx 160$~mm. Here we adopt an effective working dimension of an active volume of 60~mm~$\times$~60~mm~$\times$~2~mm. Thus the QHD has of order ${\cal L}_{xy}^2 {\cal L}_z/{\cal V} \sim 2\times10^7$ voxels, i.e. simply storing one photon per voxel would allow the storage of $N_\gamma \sim 20$ million photons. There is a very real prospect of achieving this goal.

Writing (and reading) photonic data to each voxel needs to avoid cross talk between voxels that would result in photonic information being corrupted. Ideally, during the writing (and reading) of any particular voxel, the remaining voxels would be perfectly non-interacting, that is transparent, to the signal, rephasing and control wavelengths (see Fig.~\ref{f:multiplexing}). To articulate the key ideas, we propose a system
(Fig.~\ref{f:jukebox})
where the QHD is composed of $20$ (60~mm~$\times$~60~mm~$\times$~100~$\mu$m) disks operating in a jukebox mechanism, where we assume the required crystal homogeneity and quality can be maintained over each disk. A disk is initialised into the correct spin state for storage, which simultaneously prepares $\sim 10^6$ voxels for writing. The disk is then fed into the path of the optical cavity mode. Through relative movement of the cavity mode with respect to the 2D plane of the disk, mechanically and/or through low-loss cavity/intra-cavity beam steering optics, the photons to be stored are rastered through the prepared voxels. As the voxel capacity of one disk is filled, another is fed into the cavity mode until the total QHD is loaded. Ultimately, advances in engineering spectral gradients throughout a single crystal, single voxel activation and addressing, and customised quantum control techniques should enable the jukebox to be replaced with a single continuous crystal. The jukebox mechanism is a simple way to progress towards this more elegant solution.

There are opportunities to increase the storage density by multiplexing in the frequency domain. For Eu$^{3+}$ concentrations of 1 - 1000 ppm in YSO, individual atoms exhibit narrow ($<$200 Hz) optical absorption lines (the homogeneous linewidth FWHM). Even at cryogenic temperatures, these narrow linewidths are obscured by a wider (1 - 10~GHz) inhomogeneously-broadened line due to static variations in the nanoscopic environment of each atom. While the inhomogeneous linewidth defines the range of photon wavelengths that can be stored, the bandwidth limit for hour-long storage is defined by the energy level structure of the Eu$^{3+}$ nuclear spin (frequency separations of the order of 10 - 100~MHz), which ultimately store the photonic quantum state. Using demonstrated storage parameters in our choice of material as a guide~\cite{Jobez2016, Ma2021}, it is feasible to create $(n_{\rm band} = 10)$ $\times$ 5~MHz absorption features within the inhomogeneously broadened profile within each voxel. This is possible by using optical pumping techniques to prepare frequency addressable subsets of atoms in the desired state for optical storage, so that the QHD storage capacity is increased by $n_{\rm band}$. Thus, our spectral band is $\delta\nu \approx 50$ MHz. By astronomical standards, this is {\it very} narrow ($\delta\lambda\approx 0.06$ pm at 580~nm) but is sufficient for a first on-sky demonstration of the technology when observing a bright continuum source (see \S~\ref{subs:calc}). We note that while GHz spectral bands have only been thus far demonstrated for short storage times~\cite{Saglamyurek2015,Saglamyurek2011}, inhomogeneous absorption engineering combined with advancements in spin-state storage using atoms with larger spin-state separations (eg. Er$^{3+}$) will allow broader band long-term storage in the near future.

The number of atoms available for the photon storage within each voxel is $\sim 10^{11}$. The number of active europium atoms per voxel is well-approximated by 
\begin{equation}
    n_{\rm Eu} = n_{\rm Y}  f_{\rm site} f_{\rm spec} {\cal C} {\cal V}
\end{equation}
where the number density of yttrium is $n_{\rm Y}= 1.87\times 10^{10}$ $\mu\rm m^{-3}$, the Eu concentration is ${\cal C}$, and the voxel volume is ${\cal V}$ in units of $\mu\rm m^3$. The site fraction $f_{\rm site}=0.5$ reflects the fact that there are two crystallographically distinct sites that the rare earth atom can occupy, but only one is utilised. 
The spectral fraction $f_{\rm spec}$ is the ratio of the storage bandwidth to the inhomogeneous linewidth. Specifically,
the atoms are spread out over $\Delta\nu \approx 10$ GHz in frequency space, but we only use the $n_{\rm band}\times 5$-MHz band for storage presently; thus, we set $f_{\rm spec} \approx 0.005$. 

The choice of optical storage protocol prior to mapping to the longer lived spin states will heavily influence the available write time for each voxel. While there are numerous optical quantum memory protocols, for example Electromagnetically Induced Transparency (EIT)~\cite{Lvovsky2009, Chaneliere2018}, Gradient Echo Memory (GEM)~\cite{Lvovsky2009, Campbell2016}, and Atomic Frequency Comb (AFC) techniques~\cite{Afzelius2010, Lvovsky2009}. To maximise the exposure time of the memory it will be essential to overcome the inhomogeneities in the atoms' optical frequencies. To this end, in Fig.~\ref{f:multiplexing}, we favour photon echo rephasing methods, which use two optical inversion pulses ($\pi$-pulses) to allow the writing period to be limited by the homogeneous linewidth rather than by the absorbing feature width (as is the case for GEM or AFC). For example, a photonic state absorbed by a $\delta = 5$~MHz wide atomic feature will quickly dephase as $\sim e^{-\delta t}$, whereas photon echo rephasing can allow the quantum state to be preserved on the optical transition for $e^{-2t/T_{2, \rm o}}$, where $T_{2, \rm o} = 1/(\pi \gamma_{h, \rm o})$ for a homogeneous linewidth $\gamma_{h, \rm o}$. For Eu$^{3+}$:YSO, $T_{2, \rm o} = 2.6$~ms~\cite{Equall1994} allowing a maximum exposure period of 34~$\mu$s to allow for photon recall efficiencies greater than 90\%. For simplicity we assume a exposure period of $T_{\rm expose} = 25$~$\mu$s, which provides some allowance for finite control pulse widths. These techniques include, but are not limited to, Revival of Silenced Echo (ROSE)~\cite{Dajczgewand2015}, HYbrid Photon Echo Rephasing (HYPER)~\cite{McAuslan2011a}, and Stark Echo Modulation Memory (SEMM)~\cite{Arcangeli2016}.

\subsection{Data rates}
\label{s:rates}

\begin{figure*}[htbp]
\centering
\fbox{\includegraphics[width = 0.9\textwidth]{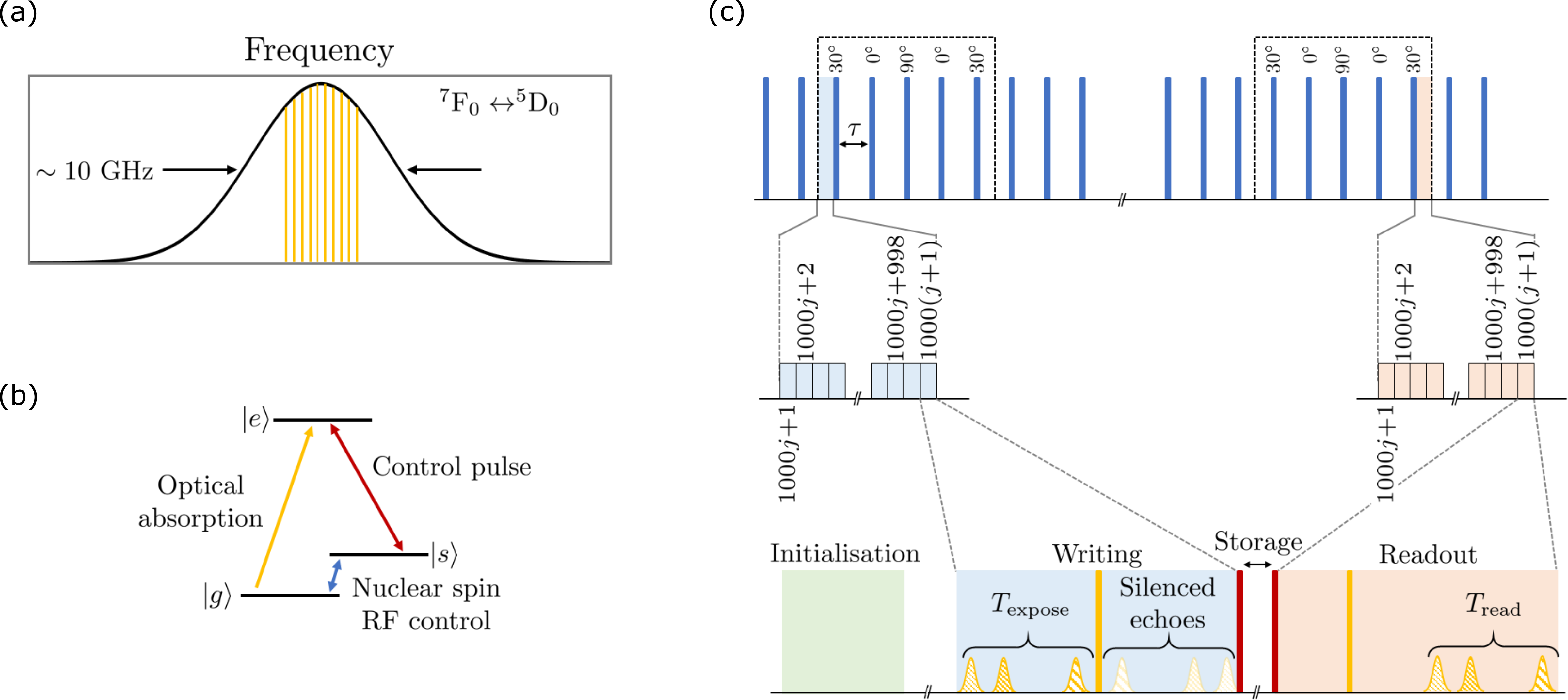}}
\caption{Writing, storing and reading photons into an atomic QHD. (a) The $^7$F$_0 \leftrightarrow$ $^5$D$_0$ optical transition of Eu$^{3+}$:YSO has an inhomogeneous linewidth of order 10~GHz. Within this 10~GHz band it is possible to identify  $n_{\rm band}\approx10$ bands of atoms that have no overlap in their optical transitions. We set the bandwidth of each absorption band to be 5~MHz. This provides a total 50~MHz bandwidth for storage. (b) The optical transition from $|g\rangle$ to $|e\rangle$ is chosen to store signal photons from the telescope. Following the cavity-assisted absorption of photons, a control pulse maps the generated atomic coherence to the nuclear spin transition $|g\rangle \leftrightarrow |s\rangle$ for long-term storage. To preserve the quantum superposition on hour-long timescales, radio frequency ($\approx10$~MHz) spin control pulses are needed to rephase slow spectral diffusion. (c) The example QHD pulse sequence discussed in the text. A series of RF inversion pulses with different phases (Knill Dynamic Decoupling - KDD$_{\rm x}$) are used to preserve an arbitrary quantum state encoded on the nuclear spin transition. The pulses are separated by time $\tau$ and we assume 1000 voxels can be written for a time $\tau / 2 = 50$~ms every 5-pulse sequence. The writing procedure uses a previously initialised voxel (e.g. disks can be initialised through optical pumping before being loaded into the optical cavity) and exposes for $T_{\rm expose}$. An optical rephasing pulse is applied and the exposure time is rephased but under conditions where the optical echoes are suppressed (silenced). The ensemble coherence is then mapped to the nuclear spin clock transition for long-term storage with an optical control pulse. The readout proceeds similarly. A control pulse maps the nuclear spin coherence back to the optical transition, which is then rephased by a second rephasing pulse. The $T_{\rm expose}$-long time window is then rephased under conditions that generate coherent emission back into the cavity. As shown, the timing of the readout is matched to the writing process such that 1000 voxels can be read in $\tau/2$.    }
\label{f:multiplexing}
\end{figure*}

After each exposure period, the first optical rephasing pulse is applied to the voxel and after a further wait time $\sim T_{\rm expose}$, a control pulse transfers the generated optical coherence to the nuclear-spin clock transition. We assume that the rephasing and control pulses are frequency multiplexed - ``multi-chromic'' - such that the optical coherence in all $n_{\rm band}$ bands are maintained and transferred simultaneously. Optical writing then commences on another voxel. This can be repeated for a time $\tau/2$ - effectively the loading time for the nuclear spin transition - where $\tau$ is the pulse separation time for the dynamic decoupling sequence. Based on previous work~\cite{Zhong2015, Ma2021}, we set $\tau = 100$~ms, which allows $10^3$ voxels to be written in series at a rate of 20~kHz, which is in the range of fast scanning optics. Optical writing is then paused to allow the full cycle of the dynamic decoupling sequence, for example for another 450~ms for KDD$_{\rm x}$~\cite{Zhong2015} (see Fig.~\ref{f:multiplexing}). This loading sequence would allow a single disk ($10^6$ voxels) to be written in 500~s and the QHD ($2\times10^7$ voxels) would take less than 3 hours, well matched to dark-time observing on an astronomical telescope.

The time taken to read the QHD is equal to the writing time (see \S~\ref{s:Read}) meaning that each voxel will have approximately the same loss due to decoherence during the storage time on the nuclear spin state. For currently demonstrated nuclear spin coherence lifetimes of $T_{2,\rm s}~=~6$~hours, $\sim$2/3 of the stored photons will be lost ($e^{-2t/T_{2,\rm s}}$ for $t = 3$~hours) during the read/write process. Even at this level of loss, the photon preservation over long distances can still dramatically outperform transmission through an optical fibre. For example, if the QHD is transported at a speed of 100~km/h for 3~hours, the overall efficiency limited by the atomic coherence lifetimes will be approximately $\eta = 0.1 = 10$~dB loss ($e^{-2t/T_{2,\rm s}}$ for $t = 6$~hours), compared to an efficiency of $\eta = 5\times10^{-5}$ (42.6 dB loss) for the lowest loss fibre transparency window or 1350~dB loss at 580~nm for the same distribution distance. In addition to improving the efficiency of distributing photonic states over long distances the QHD ability to preserve the complete quantum state will may lead to improvements in other key metrics for optical VLBI.

There are a number of extensions to our simple scheme to extend the benefit of QHD for optical VLBI. Below we consider increasing the number of photons stored by the memory by increasing the bandwidth, increasing the maximum data rate of the write and read processes, and increasing the coherence lifetime of the nuclear spin states.

Increasing the bandwidth would allow a greater fraction of the source photons to be stored, ideally approaching the bandwidths of current phase interferometry telescopes. One promising avenue to is to optimise the storage medium to allow an increase in $n_{\rm band}$, for example through controlled increases in the optical inhomogeneous linewidth, or expanding the spectral width of each band by using different isotopes, atoms, or host materials~\cite{Zhong2015}. Alternatively, frequency converting source light from outside the relevant spectral band to match the available $\delta \nu$, for example with electro-optic modulators, is another approach that could be incorporated into the protocol.

For a larger input photon rate, there are benefits to increasing the maximum data rate of the memory to reduce the decoherence loss $e^{2t/T_{2,\rm s}}$ during the time $t$ taken to write and read data. One way to achieve this is to simultaneously couple $n_{\rm cav}$ identical cavities to different voxels. For a fixed number of voxels, a data rate increase of $n_{\rm cav}$ reduces the decoherence loss during the write and read processes by a factor of $\approx e^{(n_{\rm cav}-1)/n_{\rm cav}}$. One implementation is to couple multiple macroscopic Fabry-Perot cavities to the disks in our bulk scheme. Another option is to use arrays of on-chip miniaturised quantum memories~\cite{Zhong2017a} if their efficiency and storage times can be improved to match bulk schemes. The data rate can also be improved through optimized quantum control sequences for the spin states allowing the exposure duty-cycle limit to increase from 5\% toward 50\% (reducing write/read loss by a factor of $\approx e^{0.9}$). 

Optimized quantum control should also improve the robustness of quantum-state storage through customized dynamic decoupling sequences precisely matched to desensitise the nuclear spin states from perturbations, likely extending coherence lifetimes $T_{2, \rm s}$ well beyond the current 6 hours. Increasing $T_{2, \rm s}$ reduces loss during writing, transport and reading. We also note that simply increasing the speed of transport by a factor of 3 without any other improvements maintains the QHD loss at $\sim$10~dB compared to the equivalent minimum fibre loss over 900~km of 128 dB.

We conclude with a short discussion of the dynamic range of the memory, which relates to the photonic state storage capacity {\it of a single voxel}. 
The optical exposure time (write time) for each voxel is $T_{\rm expose} = 25$~$\mu$s, which records any input photon with a spectrum contained in the $n_{\rm band}~\times$~5~MHz spectral bands. Thus, for each energy band ($h\nu_i$, where $i = 1, 2, ... n_{\rm band}$), there are $>$100 Fourier transform limited temporal modes that are accessible to the voxel ~\cite{Dajczgewand2015}. Therefore, in principle, there is the capacity to retain 1,000 photonic modes within each voxel. For weak sources or small telescope diameters, this data capacity allows us to store the light from sources despite the statistical variations in arrival times. In these cases, the memory will store vacuum states in the majority of available modes. For stronger sources and larger diameter telescopes, however, each temporal mode could contain $m = 0, 1, ..., \sim 10$ photons, ultimately limited by memory noise~\cite{Hedges2010}. Thus, in principle, there is the capacity to store $>$10,000 photons within each voxel. This storage capacity provides benefits for system testing on bright, possibly artificial, sources, or simultaneously capturing guide star photons during observations for phase matching.

\subsection{Source calculation}
\label{subs:calc}

We demonstrate that there are sufficiently bright astronomical sources to provide for a proof of principle of the idea of distributed quantum memories on existing telescopes.

The photometric scale in astronomy is tied to the zeroth magnitude star Vega. In the Eu$^{3+}$ band around 580~nm, its specific flux density is $f_{\rm Vega} = 3\times 10^{-8}$ erg cm$^{-2}$ s$^{-1}$ nm$^{-1}$. A QHD tied to a 10~m telescope sees $n_{\rm Vega} = 7\times10^8$ phot s$^{-1}$ nm$^{-1}$, assuming an overall system efficiency of 10\%.

For a total spectral band of $\Delta\nu = n_{\rm band}\times 5$-MHz $=$ 50 MHz, the photon rate drops to $n^\prime_{\rm Vega} = 4\times 10^4$ phot s$^{-1}$ $\Delta\nu^{-1}$. Thus, on average there will be 1 photon incident on the memory within the bandwidth of interest during each voxel exposure period (25~$\mu$s). That is, on average 1~photon will be written to each voxel. But given the exposure duty cycle of the QHD is 5\%, the average number of photons incident on the device $n^\prime_{\rm QM} = 2\times 10^3$ phot s$^{-1}$ $\Delta\nu^{-1}$. For the initial proof of principle, there are a dozen potential sources across the sky within a factor of 2 of Vega's brightness in the optical, and hundreds of potential sources in the near infrared. Alternatively, the count rate can be increased with a larger telescope or a storage system with a broader spectral band, faster data rate, or longer spin coherence lifetime.

Given our proposal, which is a feasible extension to existing quantum memory devices, the average data storage rate is set at 2000 voxels per second and is well matched to the incident photon flux from a star like Vega. In this example, $2\times10^7$ photons are written to the QHD over the course of approximately 3 hours. The number of photons read out after 3~hour transport time is estimated to be $\sim 2\times10^6$, limited by the coherence time (see \S 3B). This is sufficient signal to resolve a source at the limiting resolution (e.g. star's diameter), but not sufficient to resolve a source's structure (e.g. stellar limb brightening), as we show.

We can write the signal-to-noise ratio {\tt SNR} as
\begin{equation}
    {\tt SNR} = {{\sqrt{M} N \gamma}\over{\sqrt{N+\gamma^{-2}}}}
\end{equation}
for which $\gamma$ is the spatial coherence function, $N$ is the number of photons per voxel read out, and $M$ is the number of independent reads. The visibility ${\cal V}$ is defined as ${\cal V}=\vert \gamma\vert$. Detailed discussions of sensitivity calculations with interferometers are found elsewhere \citep{Dainty1979,Greenaway1979}.
To resolve the source, we seek to reach the first minimum in the visibility function with, say, ${\cal V}\approx0.1$; we expect to reach this level with ${\tt SNR} \sim 10$. Spatial imaging is much more challenging, for which ${\cal V}\approx 0.01$ in our current set up. Here we have effectively no sensitivity to structure, with  ${\tt SNR} \sim 0.1$. Even with the expected gain from a broader spectral band providing $N\sim 10$ photons per voxel read out, ${\tt SNR} \sim 1$. Spatial mapping of a source over any baseline will be challenging until quantum memories are truly broad band in their spectral response ($\Delta\nu \sim 10-100$ GHz). More gains will come from multi-telescope configurations and larger telescope apertures, but the case is already strong for us to develop QM-based networks for interferometry. More detailed calculations using the quantum Fourier transform are left to a later paper.

\subsection{Reading and mixing the stored signal \label{s:Read}}
Reading the memory is conceptually the time-reversed operation of the writing operation in which a photon is absorbed on an optical transition and the generated coherence mapped to a long coherence lifetime nuclear spin transition. Following storage, a control pulse maps the stored ensemble coherence back to the optical domain. After a wait period $T_{\rm wait} = 12.5~\mu$s in our example, the second rephasing pulse is applied enabling the ensemble to rephase and generating the emission of the stored photon or photons over $T_{\rm read} \approx T_{\rm expose}$. For this protocol, the QHD is read out at an average rate of 2000 voxels per second in a first-in-first-out sequence. Because the atomic memory records the amplitude and phase of the incident light field, the k-vectors of the photons are preserved ensuring that the photons exit the QHD in the same spatial mode in which they entered. Just as the impedance matching condition~\cite{Afzelius2010} ensured that the photons incident on the cavity are absorbed by the voxel's ensemble, it also guarantees these emitted photons couple out of the cavity with near unity efficiency in principle.

The write-store-read process for the QHD is done without knowledge of the input photonic state and without any measurement, thereby avoiding consequences of the Heisenberg Uncertainty Principle. Indeed, with sufficiently high recall efficiency~\cite{Hedges2010}, it is possible to surpass the classical bound set by the quantum no-cloning theorem, which guarantees that more information about the input light was recalled in the output light than lost or left behind in the memory. This sets a bound on how much information any observer could have as to the photon's path.  

To leverage a pair of QHDs for astronomical imaging the two streams of recalled photons must be interfered with delays appropriate to the baseline (separation between the slits). Similarly to the input to the QHD coming from the telescope, the output from the QHD can be coupled to a single mode fibre by matching the fibre and cavity spatial modes. Interference can then be measured in the spatial domain by overlapping the single-mode fibre outputs in space, or in the time domain through second order correlation measurements. Appropriate delays could be implemented through existing optical delay line technology, or by manipulating the phase of the states in the QHD prior to readout.

\section{Quantum photonics - challenges}
\label{s:challenges}

\subsection{Filtering background noise}

Any photons with frequencies outside the combined absorption features defining the $\Delta\nu = 50$~MHz QHD storage bandwidth can be a source of noise and error. Outside of the $\sim$10~GHz inhomogeneous linewidth of the $^7$F$_0 \Longleftrightarrow ^5$D$_0$ optical transition, the europium ions of interest will have isolated absorption peaks corresponding to transitions from the $^7$F$_0$ level to other 4f and 5d states (the closest in frequency being the $^7$F$_0 \Longleftrightarrow ^5$D$_1$ transition at $\sim$526~nm and the $^7$F$_0 \Longleftrightarrow ^7$F$_6$ transition at $\sim$1828~nm). A highly transmissive bandpass filter with strong rejection outside the target band will be required. 

Spectral filtering outside the 10$\times$5-MHz bands but within the $\sim$10~GHz inhomogeneous linewidth is more challenging. Photons at these frequencies need to be filtered to minimise excitation induced decoherence or direct excitation from the storage states to other excited state hyperfine levels. 

The optical cavity assists in the filtering by suppressing the majority of these photons. High frequency resolution spectral filtering with high contrast between the pass and stop bands can also be achieved with another rare-earth crystal of the same material that is optically thick (e.g. Beavan et al 2013~\cite{Beavan2013}).

\subsection{A transportable quantum hard drive \label{subs:transport}}

A key requirement for our proposal to succeed is the ability to transport the QHD without losing or corrupting the stored photonic information. To date, there are only very few demonstrations of physically transporting quantum information stored in matter-based systems. For example, architectures have been developed to shuttle atoms 100s of microns in on-chip traps for quantum computing~\cite{Bowler2012, Walther2012, Kaufmann2018}. Kaufmann et al.~\cite{Kaufmann2018} demonstrate shuttling of a $^{172}$Yb$^+$ ion over distance of 280~$\mu$m with a transport fidelity lower bound of 99.8\% and quantum state fidelity of 99.9994\%. Adding up the total distance of travel in experiments with $4\times10^3$ attempted transports, the ion would have moved of the order of 1.2~m with a quantum state fidelity of around 98\%. There have also been some initial demonstrations of physically transporting memories containing stored classical light, such as the recent work by Li et al.~\cite{Li2020}. Here the authors stored classical pulses of light in a cold ensemble of atoms in a magneto-optical trap using EIT and recalled the light with $\sim10$\% efficiency following transport of 1.2~mm. 

Matter-based QM transport distances are currently small compared to the baselines we envisage. But if the memory platform is robust against perturbations, there is no fundamental reason why quantum coherence cannot be preserved during long distance transport. The solid-state system we identify in this paper, Eu$^{3+}$:YSO, exhibits high insensitivity to perturbations and thus, considerable promise for this application. First, the clock transition is insensitive to magnetic field perturbations to first order and has low second order magnetic sensitivity, which allowed the work of Zhong et al. to be performed without any magnetic shielding~\cite{Zhong2015}. Secondly, the nuclear spin transitions are insensitive to electric fields, with Stark shifts $\leq 1$~Hz (V cm$^{-1}$)$^{-1}$~\cite{Macfarlane2014}. Finally, the hour long coherence lifetimes measured in~\cite{Zhong2015, Ma2021} already account for the sensitivity to the temperature, isotropic pressure, and acceleration fluctuations present in stationary but vibrating cryostats over both short and long time scales. Maintaining equal or improved conditions during transport is certainly a feasible engineering challenge.

We conclude this section with considerations that will impact the QHD performance 
\begin{enumerate}
    \item Maintaining an inertial environment will be important to minimise additional decoherence due to vibrations, anisotropic pressure variations, and accelerations the QHD will experience during transport. Measurements of the clock transition sensitivity to pressure - equivalently strain - and acceleration perturbations will help define the level of stabilisation required. The very low sensitivity of the $^7$F$_0 \Longleftrightarrow ^5$D$_0$ optical transition~\cite{Thorpe2011a, Thorpe2013, Zhang2020} to pressure ($\sim 200$~Hz / Pa) and acceleration ($10^{-11}$~$(\Delta f / f) / g$), suggest orders of magnitude less sensitivity for the clock transition~\cite{Zhang2020}.
    \item Minimising undesirable inhomogeneity and spectral diffusion is paramount to success, particularly given the 10$\times$ larger crystal size we consider compared to the work in Refs~\cite{Zhong2015, Ma2021}. This includes minimising inhomogeneity of the DC magnetic field creating the clock transition, the RF magnetic field used for DD control, and in the variation of the atom parameters including spin transition frequency. Optical spectral diffusion can reduce the performance of the QHD and thus needs to be carefully managed. While native optical spectral diffusion is extremely low in Eu$^{3+}$:YSO~\cite{Yano1992}, changes in external perturbations between the write and read processes can cause phase shifts, and distort the timing of the recalled signal. 
    \item To achieve hour-long coherence lifetimes, the orientation of the YSO crystal relative to the applied DC magnetic field must be maintained to within 0.004$^\circ$ to null the first order magnetic sensitivity of the Eu$^{3+}$ nuclear spin transition~\cite{Zhong2015}. This will require an extremely stable jukebox mechanism capable of maintaining the disk orientation while operating at cryogenic temperatures. This motivates the use of a single crystal QHD where voxels can be defined using diffraction limited 3D scanning of crossed laser beams.
    \item In this work, we have assumed that the various optical and RF pulses are ideal and have negligible length for simplicity. There are experimental tradeoffs between finite pulse lengths, the accuracy, and efficiency of the inversion pulses that need to be balanced to optimise the achievable number of photons that can be retrieved from the QHD~\cite{Dajczgewand2015}. We foresee that any small factor reductions from the simplified picture presented here will be offset by improved data rates, bandwidths, and optimised pulse sequences.   
    \item The fundamental limit on the storage density of a QHD is still an open research question. While alternate materials and advances in quantum memory protocols are likely to increase the photonic storage density from the $2\times10^{8}$~optical modes per cm$^{-3}$ presented here, there are fundamental limits in how far this can be extended. For example, increasing the optical depth through higher concentration samples will shorten coherence lifetimes through increased atom-atom interactions. Similarly, higher data rates will increase the heat load the QHD systems will have to dissipate to maintain a constant temperature. Experimental tests of such limitations are required to form firm bounds on the maximum achievable capacity of the QHD.
\end{enumerate}

\subsection{Summary remarks: astrophotonics and quantum photonics}

A quantum network builds on the idea of quantum teleportation where quantum information can be distributed over an optical fibre or free space link ($10-100$ km in length) if the two end nodes are connected by entangled photons. To achieve these long distances, the network must overcome fibre attenuation. One way to do this is by using a quantum repeater: a daisy-chain of QMs that break the total distance into shorter length links, and synchronise and connect successful links. The idea of using a quantum repeater network to connect separated telescopes has been discussed by several foundational papers~\cite{Gottesman2012, Khabiboulline2019, Khabiboulline2019a}, including the use of `digital' quantum memories to reduce the required rate of entangled photon generation~\cite{Khabiboulline2019}. 

Our proposal has two important differences with earlier schemes that
substantially improves the likelihood of success.
First, by physically transporting the QHD, it is possible to avoid the enormous infrastructure burden of a \emph{fixed telescope} quantum repeater network. While great strides have been made in quantum key distribution networks~\cite{Chen2021} by utilising trusted classical nodes rather than quantum-memory-enabled quantum state distribution, a fully operational quantum repeater is still a significant technical challenge. A QHD-enabled optical VLBI lowers the barrier to ultra-high resolution optical astronomy as well as greater baseline versatility, particularly if future improvements allow the telescope diameter to be reduced. This would allow large distributed networks much like VLBI radio networks today. Secondly, the memory protocol itself does not time-tag the photon arrival times as in Ref.~\cite{Khabiboulline2019}, but rather records the incoming photon signal in time over the $T_{\rm expose}$ period. In the limit of no spectral compression or stretching of the storage bandwidth (equivalently no temporal distortions to the stored signal), timestamping only need occur at the beginning of each write period with a higher accuracy than the time delay imposed by the bandwidth.

The long-term goal to realize quantum networks has spawned a new generation of photonic technologies. There is existing overlap with astrophotonics in the sense of treating light as a continuous stream of (coherent) radiation \cite{Andersen2015}.
But for quantum information processing, the discussion is often centred on light as a discrete variable (qubit). Here the photonic technologies must handle quantum states preserved in photons without destroying them, a field often referred to as {\it quantum photonics}. 
Most of our discussion is focused on the technologies underpinning quantum networks, but there are
important parallels with the drive to achieve quantum computers.
With a view to the future of optical VLBI\footnote{When we speak of `optical,' we are implicitly assuming the infrared spectrum as well. There are important applications from the optical (500 nm) out to at least the mid-IR ($\sim$10,000 nm) where astrophysical sources can be very bright in terms of photon flux. But there are very different challenges to be solved over that range; the earliest applications are expected in the 500-1500 nm window.}, this is a natural extension for the future of astrophotonics.

The clearest overlap between astrophotonics and quantum photonics as they 
exist today is the treatment of linear polarization. ESO Gravity combines the signal of all four Very Large Telescopes (VLT) as a 6-baseline interferometer over a 100m distance through remarkable photonic advances. This system 
ensures the integrity of the active laser metrology by careful control of environmental factors while maintaining the polarization state of the control signal. Related techniques are used by the MIRC-X beam combiner on the CHARA array for combining
the signal of 6 telescopes over a 300m baseline. Similar 
technologies are used in communications over classical and quantum networks. In classical networks, linear polarization is a convenient 2D degree of freedom and has been used to transmit qubits \cite{Hjelme2011}.
We anticipate rapid acceptance by the astronomical community of quantum photonics because of the potential of many enabling technologies that have yet to reach beyond the quantum community. 

\section{Acknowledgments}
JBH is funded by an ARC Laureate Fellowship and acknowledges ongoing activities in the Sydney Astrophotonics Instrumentation Labs (SAIL) that has inspired some of this work. MJS is a Chief Investigator within the Australian Research Council (ARC) Centre of Excellence for Quantum Computation and Communication Technology (CQC$^2$T, CE170100012). JGB is a Chief Investigator within the ARC Centre of Excellence for Engineered Quantum Systems (EQUS, CE170100009). We acknowledge important conversations with Rose Ahlefeldt, Julia Bryant, Chris Betters, Barnaby Norris and, in particular, with Gordon Robertson of the SUSI project. Finally, we are indebted to Celine B\"ohm for creating the Grand Challenges at the University of Sydney that provided the inspiration for this work.

\section{Disclosures}
The authors declare no conflicts of interest.

\bigskip

\bibliography{QM-interferometry.bbl}

\ifthenelse{\equal{\journalref}{aop}}{%
\section*{Author Biographies}
\begingroup
\setlength\intextsep{0pt}
\begin{minipage}[t][6.3cm][t]{1.0\textwidth} 
  \begin{wrapfigure}{L}{0.25\textwidth}
    \includegraphics[width=0.25\textwidth]{john_smith.eps}
  \end{wrapfigure}
  \noindent
  {\bfseries John Smith} received his BSc (Mathematics) in 2000 from The University of Maryland. His research interests include lasers and optics.
\end{minipage}
\begin{minipage}{1.0\textwidth}
  \begin{wrapfigure}{L}{0.25\textwidth}
    \includegraphics[width=0.25\textwidth]{alice_smith.eps}
  \end{wrapfigure}
  \noindent
  {\bfseries Alice Smith} also received her BSc (Mathematics) in 2000 from The University of Maryland. Her research interests also include lasers and optics.
\end{minipage}
\endgroup
}{}

\end{document}